**Do intelligent tutoring systems benefit K-12 students? A meta-analysis and evaluation of heterogeneity of treatment effects in the U.S.**

Walter L. Leite, walter.leite@coe.ufl.edu; University of Florida, USA

Huibin Zhang, huibinzhang1024@gmail.com, University of Tennessee Knoxville, USA

Shibani Rana, shibanigrana@gmail.com, Harvard University, USA

Yide Hao, yidehao@umich.edu, University of Michigan, Ann Arbor, USA

Amber D. Hatch, hatcha@ufl.edu, University of Florida, USA

Lingchen Kong, l.kong@ufl.edu, University of Florida, USA

Huan Kuang, hkuang2@fsu.edu, Florida State University, USA

**Do intelligent tutoring systems benefit K-12 students? A meta-analysis and evaluation of heterogeneity of treatment effects in the U.S.**

## Abstract

To expand the use of intelligent tutoring systems (ITS) in K-12 schools, it is essential to understand the conditions under which their use is most beneficial. This meta-analysis evaluated the heterogeneity of ITS effects across studies focusing on elementary, middle, and high schools in the U.S. It included 18 studies with 77 effect sizes across 11 ITS. Overall, there was a significant positive effect size of ITS on U.S. K-12 students' learning outcomes ($g$=0.271, $SE$=0.011, $p$=0.001). Furthermore, effect sizes were similar across elementary and middle schools, and for low-achieving students, but were lower in studies including rural schools. A MetaForest analysis showed that providing worked-out examples, intervention duration, intervention condition, type of learning outcome, and immediate measurement were the most important moderators of treatment effects.

**Do intelligent tutoring systems benefit K-12 students? A meta-analysis and evaluation of heterogeneity of treatment effects in the U.S.**

**1. Introduction**

Strong interest in applying artificial intelligence (AI) to education has been evidenced by multiple annual conferences on the topic and a large stream of publications (Korkmaz & Correia, 2019). Among these applications of AI, intelligent tutoring systems (ITS) have the ambitious goal of matching the benefits of one-to-one human tutoring (VanLehn, 2011), which has long been considered the most effective type of instruction (Bloom, 1984). To chase this goal, a multitude of ITS have been developed for K-12 education (Mousavinasab et al., 2018). However, the use of ITS under conditions that are not favorable may lead to wasted resources and more profound educational inequalities. As the access to computers by U.S. K-12 students has increased in recent years (Machusky & Herbert-Berger, 2022), so has the potential for students to benefit from ITS. To ensure that investments in ITS in K-12 education are both effective and strategic, it is essential to understand the conditions under which ITS use is most beneficial, but few meta-analyses have examined the heterogeneity of the effects of ITS on K-12 students in U.S. schools.

Studies of ITS have been summarized in multiple meta-analyses (Kulik & Fletcher, 2016; Ma et al., 2014; Steenbergen-Hu & Cooper, 2013, 2014; VanLehn, 2011), but the internal and external validity (Shadish, 2010; Shadish et al., 2002) of effect sizes reported for K-12 education vary greatly. Therefore, we use Campbell's validity typology (Campbell, 1957; Campbell & Stanley, 1966; Shadish et al., 2002) to evaluate ITS studies because their taxonomy of threats to validity provides a rigorous and detailed path to understanding the strengths and weaknesses of studies. The internal validity of an ITS's effects indicates the strength of the evidence that the ITS's use is causally associated with improvement in student achievement. External validity refers to whether the conclusions about the effects of ITS generalize across populations, outcomes, settings, and times (Shadish et al., 2002).

Campbell's validity typology has been extensively applied to evaluate threats to validity in single studies (Imbens, 2010; Maxwell, 2010; Shadish, 2010). However, its application to meta-analysis requires additional structure to facilitate evaluating groups of studies. To provide the needed structure, we applied the MUTOS framework (Becker, 2017) to group studies by dimension and link these dimensions to threats to validity. Specifically, the MUTOS framework, extending Cronbach's (1982) UTOS framework, refers to five dimensions of studies: methods (M), units (U), treatments (T), observing operations (O), and setting (S). An example of the M dimension includes study design choices such as whether the study performed random assignment at level 1 (i.e., students), 2 (i.e., teachers/classrooms) or 3 (i.e., schools); the U dimension includes characteristics of samples, such as whether the study targeted low-achieving or a general population of students; the T dimension refers to characteristics of the ITS such whether it provides just-in-time or on-demand hints and characteristics of the treatment implementation such as its duration; the O dimension refers to measurement, such as whether the outcome was measured by a researcher-developed test or a standardized test; and the S dimension refers to settings such as the location of participants. In the MUTOS framework, the M dimension provides evidence of internal validity, while the U, T, O, and S dimensions allow examination of external validity.

The current study addresses the scarcity of information about heterogeneity of the effects of ITS in the U.S. K-12 educational system by responding to the following research questions:

(1) What is the effect of ITS on student academic achievement in U.S. K-12 schools? (2) Do the effects of ITS depend on whether the study provides strong evidence of internal validity? (3) Do the effects of ITS have strong evidence of external validity? To address the second question, we evaluate study characteristics in the M dimension, the threats to internal validity that emerge from these characteristics, and whether the study meets the What Works Clearinghouse (WWC) standards with or without reservations (U.S. Department of Education et al., 2022). For the third question, we examine the extent to which the effect sizes vary across the U, T, O, and S dimensions.

The study is organized as follows: first, we review the definition and characteristics of ITS, and the effect sizes estimated in previous meta-analyses. We then review the evidence of internal validity and external validity of the effects reported in previous meta-analyses, using the MUTOS framework to organize the evidence. Based on this review, we present the hypotheses evaluated in the current study. This is followed by a description of the study's methods and results. The discussion and conclusion section contrasts the results of the current work with those of previous meta-analyses, reviews limitations, and outlines directions for future research.

## 2. Theoretical Framework

### 2.1. Intelligent Tutoring Systems

ITS is software that interacts with students on a turn-by-turn basis, personalizing each turn to the individual student based on a model of student learning (D'Mello & Graesser, 2024; VanLehn, 2011). ITS must keep track of the status of primary and secondary student outcomes to provide personalized support, which is accomplished through learner models. These include probabilistic models for academic achievement, such as Bayesian Knowledge Tracing (BKT;

(Bulut et al., 2023; Šarić-Grgić et al., 2024), but may also include other cognitive states, such as affect, engagement, and attention (D'Mello & Graesser, 2024).

ITS can have different degrees of granularity of interaction with the student, such as interaction at each step or sub-step (VanLehn, 2011). The interaction may include various learning supports, such as instructional content, personalized interactive practice, hints, feedback, and encouragement. Instructional content may include text, videos, multimedia presentations, and worked-out examples. These serve as the foundational material upon which learners build their understanding. Worked-out examples are step-by-step demonstrations of how to solve a problem or perform a task. They help learners understand the process and reasoning behind solutions. Hints and feedback provide guidance and corrective information during learning activities, helping learners stay on track and understand mistakes. Hints can be just-in-time hints, which are provided automatically by the ITS upon detecting that a student is struggling with a problem. On-demand hints are those that the student may request at any time during the problem-solving process.

**2.2. Effects of ITS on Student Achievement**

VanLehn (2011) conducted a meta-analysis of ITS and found positive effects of different ITS granularities compared to no tutoring. Specifically, step-based ITS showed an effect size of $d = 0.76$, sub-step-based ITS had an effect size of $d = 0.40$, and answer-based ITS had an effect size of $d = 0.31$. Ma et al. (2014) reported positive effects of ITS compared to large-group human instruction ($g = 0.44$), computer-based instruction ($g = 0.577$), and individual textbook instruction ($g = 0.36$). These results indicate that ITS can be beneficial across various instructional settings. Kulik and Fletcher (2016) estimated an effect size of $g = 0.62$ for ITS, further supporting the effectiveness of ITS in improving educational outcomes. These previous findings lead to our first hypothesis:

*Hypothesis 1: ITS use in U.S. K-12 schools has a positive effect on student achievement.*

## 2.3. Internal Validity of Effects Reported in Previous Meta-Analyses

Internal validity depends on the characteristics of the study design and analysis used to estimate the treatment effect, such as the type of study design (e.g., experimental, quasi-experimental, correlational). Experimental designs provide effect estimates with the strongest internal validity, but they may be compromised by attrition (Shadish et al., 2002). From existing meta-analyses of ITS, only Ma et al. (2014) reported attrition, but their study included all study designs, which may suffer from other threats to internal validity, such as selection.

The strength of internal validity of the effects of ITS reported in previous meta-analyses varies across factors in the M dimension of the MUTOS framework, such as type of design, type of random assignment, and attrition level. We focused on studies that meet the WWC standards (U.S. Department of Education et al., 2022). These are rigorous and comprehensive criteria to indicate the extent to which interventions are causally linked to educational outcomes. Experimental designs with low differential attrition meet the WWC standards without reservations, providing the strongest evidence of causality. Experimental designs with high differential attrition meet the WWC standards with reservations. However, previous meta-analyses also included weaker designs, such as quasi-experimental designs and correlational designs that do not meet the WWC standards. While quasi-experimental designs may meet WWC standards with reservations if they demonstrate baseline equivalence, correlational designs do not.

VanLehn (2011) only included experimental studies published with individual-based treatment assignments, excluding cluster-randomized trials (CRT), where treatment assignment is at level 2 (e.g., teacher or classroom) or level 3 (e.g., schools). However, CRT studies are as rigorous as individual-level experimental designs and meet WWC standards without reservations. Because of clustering effects, CRTs require a larger sample size than individual-level experimental designs to achieve the same power level (Spybrook et al., 2014). Steenbergen-Hu and Cooper (2013) used the WWC standards to inform the selection of studies, including only experimental designs that met the WWC criteria without reservations, or quasi-experimental studies that met the criteria with reservations. Kulik and Fletcher (2016) also included both experimental and quasi-experimental studies. They did not refer to WWC standards for selecting studies. However, they required that the standardized difference in baseline variables between treatment and control groups not exceed 0.5 standard deviations, which is a very lenient criterion. The WWC standards are much stricter, requiring a standardized difference of 0.05 for adequate baseline equivalence or 0.25 if covariate adjustment is performed in the outcome model. In contrast, Ma et al. (2014) did not restrict the sample to high-quality experimental studies, including all study designs.

The WWC standards require that differential attrition, which is the difference in attrition between treatment and control groups, not exceed a certain level for the experimental study to be rated as meeting the WWC standards without reservations. Rather than a single cutoff of differential attrition, the WWC Procedures and Standards Handbook V5.0 (U.S. Department of Education et al., 2022) provides a table showing the level of differential attrition resulting in unacceptable levels of potential bias for overall attrition between 0% and 45%. Furthermore, for each percentage of overall attrition, there is a cautious boundary and an optimistic boundary of differential attrition. The WWC recommends using the optimistic boundary if the attrition is unlikely to be related to the intervention and the cautionary boundary otherwise. In this meta-analysis, we decided to use the optimistic boundary.

Previous meta-analyses provide limited evidence about internal validity at the level of WWC standards because they did not code whether studies reported attrition. The only exception is Ma et al. (2014), which found that 21.5% of studies reported some attrition, 33.6% reported no attrition, and 44.9% did not. From Ma et al.'s (2014) results, it is problematic that the studies that did not report attrition had a larger weighted mean effect size (i.e., $g = 0.48$) than studies with attrition ($g = 0.29$) and no attrition ($g = 0.39$) because it may indicate selectivity in reporting attrition. However, Ma et al. (2014) included experimental, quasi-experimental, and correlational, designs. Therefore, other internal validity threats, such as selection and history (Shadish et al., 2002), may have affected the estimates reported. These previous findings concerning the impact of attrition on evaluations of ITS lead to this study's second hypothesis:

*Hypothesis 2: The effect of ITS on student achievement in US K-12 schools is larger in studies that meet the WWC standards without reservation than in studies that meet the WWC standards with reservation.*

## 2.4. External Validity of Effects Reported in Previous Meta-Analyses

Previous meta-analyses of ITS examined different aspects of the U, T, O, and S dimensions of the MUTOS framework. When comparing effect sizes within and between previous meta-analyses, and in the current study, we considered effect sizes with less than a 0.05 standard deviation difference to be similar, because this difference is unlikely to be of practical importance. In the Units (U) dimension, previous studies have examined grade level, previous knowledge, and disadvantaged targeting. Ma et al. (2014) reported larger effects for elementary ($g$ = 0.31) than middle ($g$ = 0.41) and high schools ($g$ = 0.41). Kulik and Fletcher (2016) reported a Glass D of 0.44 for elementary and high school students combined. In contrast, for mathematics ITS in the U.S., Steenbergen-Hu and Cooper (2013) reported a large variability of effects, with $g$ = 0.41 for elementary, $g$ = 0.09 for middle, and $g$ = -0.09 for high school using a fixed effects model.

Variation in effect sizes was also associated with previous knowledge of students. Ma et al. (2014) reported lower effects for studies focusing on participants with low domain knowledge ($g$ = 0.38) and medium domain knowledge ($g$ = 0.28) as compared to studies with participants of varying domain knowledge ($g$ = 0.48), but all effects were positive. In contrast, Steenbergen-Hu and Cooper (2013) reported a negative effect of $g$ = -0.18 for studies with samples of low-achieving K-12 students and $g$ = 0.04 with a general sample of K-12 students.

In the Treatment (T) dimension, meta-analyses of ITS have examined the type of control group, duration of intervention, and intervention support type. The types of control groups used in ITS studies vary substantially. VanLehn (2011) compared ITS to no tutoring conditions, which primarily involved students reading text individually and solving problems without feedback, rather than regular classroom instruction. Steenbergen-Hu and Cooper (2013) included in their meta-analysis control groups that received classroom instruction, one-to-one tutoring, or homework. This diversity in control groups can influence the estimated effects of ITS (Steenbergen-Hu & Cooper, 2013). Kulik and Fletcher (2016) restricted their control group to classroom instruction, which may provide a more consistent basis for comparison across studies (Kulik & Fletcher, 2016).

Shorter interventions showed larger effect sizes than longer interventions in previous meta-analyses, but classifications of duration were not consistent. Kulik and Fletcher (2016) classified the study duration into two groups: up to 7 weeks and 8 weeks or more. They found that shorter studies had effect sizes of 0.13 standard deviations higher than longer studies. Steenbergen-Hu and Cooper (2013) classified studies of ITS into three categories: short-term studies had $g = 0.52$, one-semester studies had $g = 0.06$, and studies lasting one year or longer had $g = -0.02$.

    Previous meta-analyses of ITS have provided limited evaluation of the variation of effect sizes across the types of support provided by ITS, such as on-demand hints, just-in-time hints, feedback/explanations, motivation support, reading materials, and video sequencing. Van Lehn (2011) compared ITS with respect to the granularity of support, which he classified into three types: (1) Answer-based ITS provide support after the student completes a learning task (i.e., answers a problem); (2) Step-based ITS provides support at each of the natural steps that are necessary for the completion of the learning task; and (3) Sub-step-based ITS provide support at a granularity that is finer than at every natural step of the learning-task completion. In studies using no tutoring as a control group, Van Lehn (2011) reported effect sizes of $g = 0.31$ for answer-based, $g = 0.76$ for step-based, and $g = 0.40$ for sub-step-based ITS. In contrast, Kulik and Fletcher (2016) found similar effect sizes of $g = 0.60$ and $g = 0.63$ for step-based and sub-step-based support, respectively, using regular instruction as the control group. However, the specific type of support provided by ITS with each granularity can vary widely. For example, a step-based ITS may provide on-demand and just-in-time hints, while another one may provide motivation support, on-demand hints, but no just-in-time hints. Therefore, in the current study, we complement the knowledge about the effects of granularity types provided in previous studies by examining the treatment effect heterogeneity across specific types of support provided by ITS.

Previous meta-analyses examined variation across factors in the operations (O) dimension, including the type of measure, subject of learning, type of question, and measurement time. Variation in effect sizes due to the type of measure was inconsistent across meta-analyses. For example, Kulik estimated an effect size of 0.73 with local measures and only 0.13 with standardized measures. For reading comprehension ITS, Xu et al. (2019) reported that the researcher-developed measures produced effect sizes that were, on average, 0.734 larger than standardized measures. However, Ma et al. (2014) reported very similar effect sizes of $g=0.41$ ($p < .05$) and $g=0.42$ ($p < .05$) for both types of measures.

Ma et al. (2014) identified similar effects for mathematics ITS ($g = 0.35, p < .05$) and language and literacy ITS ($g = 0.35, p < .05$), but their meta-analysis included samples from elementary school to post-secondary. For K-12 students, Xu et al. (2019) estimated an effect of $g = 0.60$ ($p < .05$) on reading comprehension, while Steenberger-Hu and Cooper (2013) estimated an effect of $g = 0.09$ ($p > .05$) on mathematics achievement.

The review of results of previous meta-analyses of ITS presented above indicates large differences in effect sizes across factors in the U, T, O, and S dimensions. Furthermore, previous meta-analyses tended to produce inconsistent results, with some showing similar effects across levels of a factor while others showing large differences. However, these marginal effects of factors in the U, T, O, and S dimensions may mask complex interactions between factors. Previous meta-analyses assumed no interactions, possibly due to the difficulty in estimating interactions with meta-regression. To circumvent this limitation, we focused on the evaluation of variable importance using meta-forests (Van Lissa, 2020), which is a non-parametric machine-learning approach. That is, considering all possible higher-order interactions that a factor in the U, T, O, and S domains may have, what is the importance of each factor in determining the effect size of ITS? Informed by differences in effect sizes reviewed above, the third hypothesis in this study is:

*Hypothesis 3: Factors in the T dimension will have higher importance in explaining variation in effect sizes than factors in the U, O, and S dimensions.*

## 3. Methods

### 3.1. Search

The systematic review underwent three stages: literature search, article screening, and coding. For the literature search, we conducted electronic bibliographic searches from the following databases: Learntechlib, ERIC, PsycInfo, Academic Search Premier, IEEE Xplore Digital Library, ACM Digital Library, and Proquest Dissertation and Theses for publications between 2011 and 2021. These databases were chosen because they cover educational research published in journal articles, conference proceedings, theses, and dissertations. The search string was created by consulting the keywords used in previous meta-analyses of ITS (Kulik & Fletcher, 2016; Ma et al., 2014; Steenbergen-Hu & Cooper, 2013, 2014; VanLehn, 2011) and using the wild card character to increase matches. The specific search string used is shown in Appendix 1.

### 3.2. Screening

The study screening was performed using the Covidence.org platform, which was developed by the Cochrane group and is widely used for systematic reviews in the health sciences (Veritas Health Innovation, 2023). We conducted abstract screening and full text screening with the inclusion and exclusion criteria shown in Appendix 2. We focused on the most rigorous designs, which must meet the WWC standards without or with reservations. In each stage, every article was independently screened by two trained reviewers. Training consisted of having all reviewers rate a sample of 20 abstracts independently, discussing and analyzing all discrepancies until all reviewers reached consensus (Polanin et al., 2019). The articles that passed the full-text screening were used for a forward and backward citation search. The new articles identified through forward and backward citation searches underwent the same two-stage screening process again. Then, all articles that passed the two-stage screening were included in the coding phase.

**3.3. Data Extraction**

We coded the data from each study using the MUTOS framework. The elements coded using the MUTOS framework were chosen according to our research objectives and previous meta-analyses of ITS. These elements are shown in Appendix 3. Several moderator variables (e.g., gender and ethnicity) had substantial amounts of missing data (i.e., exceeding 50%). Therefore, we excluded those covariates from the subsequent analyses to avoid introducing bias or reducing model stability.

We utilized the standardized mean difference (SMD) between intervention and control groups as the common metric for effect size comparison. In addition, we used Hedges' *g* to adjust SMD to reduce small sample bias (Hedges, 1981). If the mean and standard deviations of learning outcomes were unavailable, we tried to identify a t-test, or F-test results to calculate SMD (see Appendix 4). If the treatment assignment occurred at the cluster level i.e., teacher/ classroom or school), we adjusted the effect size variances for clustering (Taylor et al., 2022).

**3.4. Estimation of Effect of ITS on Student Achievement**

We estimated the effect of ITS by using a multivariate multilevel meta-regression model (Pustejovsky & Tipton, 2022) with random effects of studies and fixed effects of the specific ITS used in the studies. We decided to include ITS names as a fixed effect rather than a random effect because fixed effects do not require any distributional assumptions (Allison, 2009; Gardiner et al., 2009). Also, the fixed effect of ITS names controls for all confounding at the ITS level (Allison, 2009), eliminating the need to include ITS-level covariates. The fixed effect of ITS names was included in the model by creating an effect-coded indicator for each ITS name. We used weighted least squares (WLS) to estimate the meta-regression model parameters since it uses weighting matrices to improve accuracy of the estimation (Pustejovsky & Tipton, 2022). The model was estimated using the R package *metafor* (Viechtbauer, 2010). To adjust the regression coefficients' standard error and degrees of freedom for hypothesis tests, we applied robust variance estimation (Hedges et al., 2010) implemented with the R package *clubSandwich* (Pustejovsky, 2025). This procedure provided cluster-robust standard errors that accounted for the dependence of effects within studies. We also created a funnel plot based on the observed effect sizes and their standard errors. We provide the R code and data used in this article in an Open Science Framework site (Anonymous Contributors, 2025).

### 3.4. Analysis of Internal Validity

We evaluated each study's internal validity by examining the type of design, assignment level, levels of clustering, type of assignment (i.e., simple or blocked), and differential attrition. These are characteristics of the M dimension of the MUTOS framework. The type of design refers to whether the study is experimental or quasi-experimental. To evaluate the type of random assignment, we examined whether assignment was at level 1 (i.e., students), level 2 (i.e., teachers/classrooms), or level 3 (i.e., schools). The levels of clustering refer to whether students are in multiple clusters. For example, a study including students in multiple classrooms and schools has level 3 clustering, but if all classrooms are in a single school, the study has level 2 clustering. Also, the design has level 2 clustering if each school has a single classroom. Blocking consisted of separating units into blocks prior to random assignment and assigning participants to conditions within each block. For example, in a 3-level cluster-randomized trial with assignment at level 3, schools could be first pair-matched by student demographics, and then one school would be randomly assigned for treatment in each pair. Blocking is performed to decrease potential imbalance between covariate distributions across treatment and control groups, and to increase statistical power.

To evaluate differential attrition, we separated studies into those that reported and did not report attrition. For those reporting attrition, we identified whether separate attrition rates were provided for treatment and control groups. Then, to identify whether the study's reported level of attrition met the low attrition conservative standards of the WWC, we calculated both the overall attrition rate and differential attrition rate between groups. By comparing those values to the WWC attrition standard plot (U.S. Department of Education et al., 2022), we classified each study as having either a high or a low level of attrition. The M dimension also includes the type of publication (i.e., article, dissertation), but this study characteristic is not related to internal validity. We estimated effect sizes for studies with each component of the M dimension using a multivariate multilevel meta-regression model fit to the data of the subset of studies with each level of a component (Pustejovsky & Tipton, 2022), including random effects of studies and fixed effects of ITS names (see more details in the next section).

### 3.5. Analysis of External Validity

We examined the external validity of studies by evaluating whether there was heterogeneity of treatment effects across the moderators in the U, T, O, and S dimensions. This evaluation was performed by calculating separate effect sizes and confidence intervals for the levels of the moderators. This was done by creating a separate dataset for each level of a moderator and fitting a multivariate multilevel meta-regression model with random effects of studies. Fixed effects of ITS names included as effect-coded covariates. Thus, the effect size of each level of the moderator is estimated as the intercept of the meta-regression model. By fitting a separate meta-regression model to each level of the moderator, this allowed the fixed effects of ITS names to vary across levels of the moderator, implicitly accounting for any interactions between the moderators and ITS names. For some moderator levels, each study used a different ITS, making it impossible to separate the random study effects from the fixed ITS effects. In these cases, we only included the random study effects, as indicated in the table notes in the results section.

We also fit a MetaForest model to the entire dataset to identify important moderators that substantially contribute to the model's predictions. MetaForest is a random forest model (Breiman, 2001) for meta-analysis. This machine learning model is based on recursive partitioning (Strobl et al., 2009) and has the advantage of automatically detecting complex interactions and nonlinear effects of moderators. The MetaForest model included all moderators coded under the U, T, O, and S dimensions. A total of 15 moderators were included in the initial MetaForest model, from which 11 were dummy indicators. Details of permutation importance (Altmann et al., 2010) calculation and hyperparameter tuning are provided in Appendix 5.

### 3.6. Publication Bias and Sensitivity Analysis

Publication bias means that studies with significant results are more likely to be published than studies with non-significant effect sizes, creating a high possibility that the meta-analysis might overestimate the true effect size. We calculated Rosenthal and Orwin's fail-safe N to check how many unpublished studies would change the significance of the conclusion (Orwin, 1983; Rosenthal, 1979).

## 4. Results

### 4.1. Search Results

Figure 1. PRISMA of the Systematic Review Process

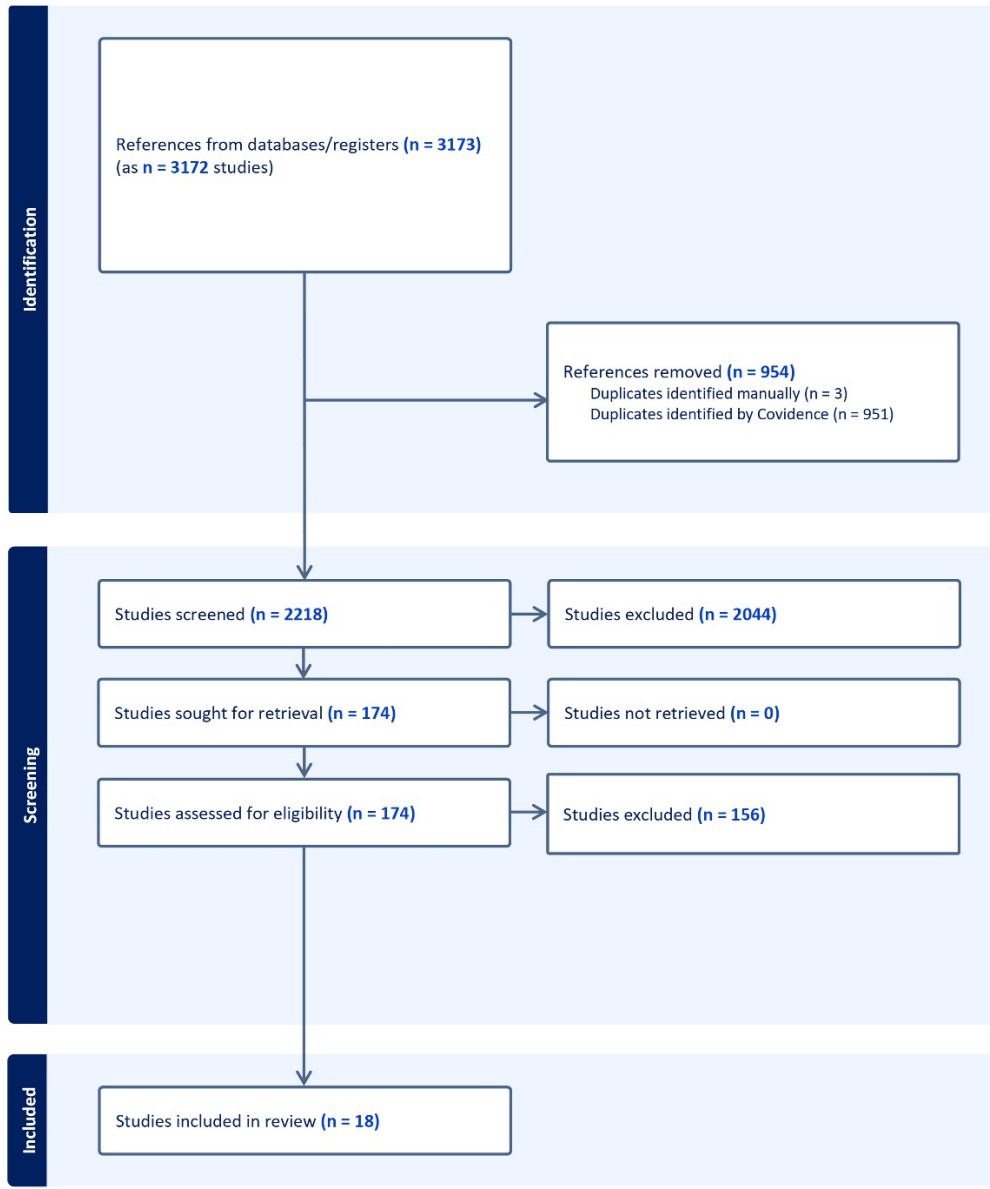

Figure 1 shows the PRISMA plot of the systematic review process. The literature search resulted in 2218 studies after removal of duplicates. The abstract screening stage excluded 2044 articles and moved 174 articles to the full-text screening stage. We identified 18 experimental studies that met all inclusion criteria, resulting in 77 effect sizes from 11 ITS (see Table 1). The systematic review did not identify any quasi-experimental studies that would meet the WWC standards with reservations[1].

**Table 1**. Frequencies of effects by ITS

| ITS | Studies | Effect Sizes | Percentage |
| --- | --- | --- | --- |
| The vocabulary intelligent tutoring system | (Baker et al., 2021) | 6 | 7.79 |
| Assessment and Learning in Knowledge Spaces (ALEKS) | (Craig et al., 2013; Hu et al., 2012; Huang et al., 2013; Huang et al., 2016) | 6 | 7.79 |
| AnimalWatch (AW) | (Niaki et al., 2019) | 1 | 1.30 |
| Cognitive Tutor Algebra I (CTAI) | (Pane et al., 2016) | 2 | 2.60 |
| MathConceptz | (Affriie-Adams, 2020) | 1 | 1.30 |
| MATHia | (Mariano, 2015) | 6 | 7.79 |
| Reasoning Mind's Grade 5 Common Core Curriculum (RM-CC5) | (Shechtman et al., 2019) | 1 | 1.30 |

---

[1] Huang et al. (2016) referred to their study as quasi-experimental, but their description of the random assignment procedure indicates that their study was experimental.

| | | | |
|---|---|---|---|
| Strategy instruction on the Web for Spanish speaking English learners (SWELL) | (Wijekumar et al., 2018) | 10 | 12.99 |
| Intelligent Tutoring System for the Structure Strategy (ITSS) | (Wijekumar et al., 2020; Wijekumar et al., 2017; Wijekumar et al., 2014; Wijekumar et al., 2022; Wijekumar et al., 2012) | 39 | 50.65 |
| Wayang Outpost | (Arroyo et al., 2011) | 4 | 5.19 |
| Woot Math Adaptive Learning (WMAL) | (Bush, 2021) | 1 | 1.30 |
| Total | 18 | 77 | 100 |

## 4.2. Effects of ITS on Student Achievement

Overall, there was a statistically significant positive effect of ITS on U.S. K-12 students' learning outcomes ($g=0.271$, $SE=0.011$ $p=0.001$) after accounting for the random effect of studies and the fixed effect of ITS names.

## 4.3. Internal Validity

Because the studies identified for this meta-analysis were all experimental, they potentially provided the highest level of evidence of internal validity. However, their internal validity may be threatened by attrition. Nine out of the 18 studies reported total attrition, but only eight reported separately for treatment and control groups. Therefore, it was only possible to calculate differential attrition for eight studies. Among these, three demonstrated high differential attrition, while five showed low differential attrition. Studies with high differential attrition ($g=0.261$) exhibited slightly higher effect sizes than those with low differential attrition ($g=0.212$) or no reported differential attrition ($g=0.251$; see Table 2). However, only the effect size estimated from studies with no reported attrition was statistically significant.

**Table 2.** ITS Intervention Effect Sizes by Differential Attrition Levels

| Attrition Level | K | $g$ (SE) | 95% CI | $p$-value |
|---|---|---|---|---|
| High | 12 | 0.261 (0.050) | (-0.081, 0.603) | 0.072* |
| Low | 11 | 0.212 (0.083) | (-0.045, 0.468) | 0.080* |
| Not reported | 54 | 0.251 (0.029) | (0.150, 0.352) | 0.006 |

*Note.* K = number of effect sizes, * indicates fixed ITS effects not included due to every study having a different ITS.

### 4.3.1. M Dimension

For the M dimension, we coded five variables: type of study design, assignment level, levels of clustering, simple or blocked design, and type of publication. As explained earlier, the type of study design included only experimental studies because no quasi-experimental study met our inclusion criteria. Table 3 shows the effect sizes across levels of M dimension moderators, accounting for the random effect of studies, as well as fixed effects of ITS names when applicable. Studies with level 2 assignment produced the smallest effect size ($g=0.269$), followed by level 1 ($g=0.331$), and level 3 ($g=0.456$) assignments. However, only the effect size of studies with level 2 assignment was statistically significant. The effect size with three levels of clustering ($g=0.253$) was similar to two levels of clustering ($g=0.238$), but only the former was statistically significant. Studies with a simple design (g=0.311) and a blocked design (g=0.271) produced similar effect sizes.

Dissertations or theses showed the largest effect size (g=0.625), while journals (g=0.190) and conference proceedings (g=0.139) had smaller effect sizes. However, only the effect size from journal articles was statistically significant. Although publication type is in the M dimension, it has no theoretical relationship to internal validity but may be related to publication bias (see section 4.4.7).

**Table 3.** ITS Intervention Effect Sizes for M Dimension

| Moderator | K | g (SE) | 95% CI | p-value |
|---|---|---|---|---|
| **Assignment Level** | | | | |
| Level 3 | 7 | 0.456 (0.399) | (-1.424 2.337) | 0.380* |
| Level 2 | 68 | 0.269 (0.010) | (0.230, 0.308) | <0.001 |
| Level 1 | 2 | 0.331 (0.408) | (-4.853, 5.515) | 0.566* |
| **Levels of Clustering** | | | | |
| 2 | 11 | 0.238 (0.136) | (-0.438, 0.914) | 0.240* |
| 3 | 66 | 0.253 (0.014) | (0.210, 0.297) | <0.001 |
| **Simple or Blocked Design** | | | | |
| Simple | 52 | 0.311(0.035) | (0.186, 0.435) | 0.006 |
| Blocked | 24 | 0.271 (0.098) | (-0.068, 0.609) | 0.081* |
| **Type of Publication** | | | | |
| Conference proceeding | 2 | 0.139 (0.063) | (-0.663, 0.941) | 0.271* |
| Dissertation or thesis | 7 | 0.625 (0.108) | (-0.704 1.954) | 0.107* |
| Journal article | 68 | 0.190 (0.016) | (0.138, 0.241) | 0.001 |

*Note.* K = number of effect sizes, * indicates Fixed ITS effects not included due to every study having a different ITS.

### 4.4. External Validity

### 4.4.1. U Dimension

The U dimension included three variables: grade level, student disadvantage targeting, and prior domain knowledge level. Three additional variables (i.e., group size, gender, and ethnicity) were excluded from the analysis due to more than 50% missing values. Table 4 provides the estimated effect sizes for each moderator level in the U dimension. The effect size for middle school students was similar ($g=0.256$) to the effect size for elementary school students ($g=0.266$). The effect size estimated from studies including both middle and high school students was higher, but not statistically significant ($g=0.335$). Also, the effect sizes were similar across studies targeting disadvantaged groups: English Learner students ($g=0.266$), students in poverty ($g=0.303$), and low achievers ($g=0.270$). Only the effect size for low achievers was statistically significant. Studies that did not target specific disadvantaged groups (i.e., the general student population) had a smaller but statistically significant effect size ($g=0.216$). Studies targeting students with low prior knowledge had a similar effect size ($g=0.234$) to those without this focus ($g=0.275$), but only the latter was statistically significant.

**Table 4**. ITS Intervention Effect Sizes for U Dimension

| Moderator | K | g (SE) | 95% CI | p-value |
|---|---|---|---|---|
| **Grade Level** | | | | |
| Elementary | 44 | 0.266 (0.033) | (0.170, 00363) | 0.002 |
| Middle | 30 | 0.256 (0.011) | (0.209, 0303) | 0.002 |
| Middle and High | 3 | 0.335 (0.353) | (-4.156, 4.826) | 0.517* |
| **Student disadvantage targeting status** | | | | |
| English language learners | 16 | 0.266 (0.070) | (-0.627, 1.158) | 0.165* |
| Low achievers | 21 | 0.270 (0.036) | (0.063, 0.477) | 0.034* |
| Students in poverty | 12 | 0.303 (0.121) | (-0.372, 0.977) | 0.161* |
| No targeting (general population of students) | 48 | 0.216 (0.017) | (0.151, 0.281) | 0.003 |
| **Low prior domain knowledge level** | | | | |
| Yes | 11 | 0.234 (0.079) | (-0.768, 1.236) | 0.207* |
| No | 66 | 0.275 (0.013) | (0.227, 0.322) | <0.001 |

*Note.* K = number of effect sizes, * indicates Fixed ITS effects not included due to every study having a different ITS.

### 4.4.3. T Dimension

Table 5 shows that the ITS provided eight different types of support to enhance learning. Feedback/explanations and question sequences were included in all 11 ITS. Video sequence was the least common support type, reported in only two ITS. The effect sizes for each moderator in the T dimension are shown in Table 6. The effect sizes were similar for $g=0.337$ for just-in-time hints, worked-out examples ($g=0.324$), and on-demand hints ($g=0.316$). In contrast, effect sizes were lower for motivation support ($g=0.265$), reading materials ($g=0.247$), and video sequences ($g=0.243$). The effect sizes estimated for each type of support were statistically significant except for just-in-time hints and video sequences. All 11 ITS provided feedback/explanations, so a separate effect size for this group was not computed.

**Table 5.** Frequency of Support Provided by ITS

| Support | Count | Percentage |
|---|---|---|
| Feedback/explanations | 11 | 100 |
| Question sequence | 11 | 100 |
| Just-in-time hints | 4 | 36.4 |
| On-demand hints | 5 | 45.4 |
| Motivation support | 4 | 36.4 |
| Worked-out example | 6 | 54.6 |
| Reading materials | 4 | 36.4 |
| Video sequence | 2 | 18.2 |

*Note.* Count is the number of ITS offering the type of support

Interventions lasting more than two weeks but up to three months ($g=0.449$) had the largest effects, followed by those lasting up to two weeks ($g=0.334$), then those lasting more than three months but up to six months ($g=0.225$). Interventions lasting more than six months had the smallest ($g=0.110$) effect size. Only interventions between three and six months had a statistically significant effect size.

Studies using ITS as the main instructional method ($g=0.294$) had similar effect size as those using ITS as homework ($g=0.317$), and both were statistically significant. Studies using

ITS as separate activities (*g*=0.235) showed a nonsignificant smaller effect. Studies using ITS in after-school settings (*g*=0.014) resulted in the smallest, non-significant effect.

Studies where different instructors taught the treatment and control groups showed a significant moderate effect size (*g*=0.282). The single study with the same instructor for both groups yielded a smaller, non-significant effect size (*g*=0.171).

**Table 6**. ITS Intervention Effect Sizes for T Dimension

| Moderator | K | *g* (SE) | 95% CI | *p*-value |
|---|---|---|---|---|
| **ITS Support Type** | | | | |
| Just-in-time hints | 18 | 0.337 (0.142) | (-0.147, 0.820) | 0.108 * |
| On-demand hints | 37 | 0.316 (0.041) | (0.085, 0.547) | 0.031 |
| Motivation Support | 23 | 0.265(0.027) | (0.102, 0.428) | 0.024 |
| Worked-out example | 59 | 0.324 (0.020) | (0.176, 0.472) | 0.018 |
| Reading materials | 40 | 0.247 (0.049) | (0.054, 0.440) | 0.031 |
| Video sequence | 2 | 0.243 (0.110) | (-1.155 1.640) | 0.271* |
| **Duration of Intervention** | | | | |
| Up to two weeks | 12 | 0.334 (0.139) | (-0.162, 0.830) | 0.112* |
| More than two weeks but up to three months | 12 | 0.449 (0.269) | (-0.497, 1.395) | 0.209* |
| More than three months but up to six months | 38 | 0.225(0.009) | (0.171, 0279) | 0.006 |
| More than six months | 15 | 0.110 (0.065) | (-0.077, 0.297) | 0.173* |

**Intervention Condition**

| | K | Estimate (SE) | 95% CI | p |
|---|---|---|---|---|
| Main Instructional Method | 45 | 0.294 (0.019) | (0.241, 0.348) | <0.001 |
| Separate Activity | 16 | 0.235 (0.079) | (-0.189, 0.660) | 0.122* |
| After School | 6 | 0.014 (0.036) | (-0.110, 0.138) | 0.723[+] |
| Individual Homework | 10 | 0.317 (0.049) | (0.205, 0.428) | <0.001[+] |

**Instructor type**

| | K | Estimate (SE) | 95% CI | p |
|---|---|---|---|---|
| Different instructors for treatment and control groups | 73 | 0.282 (0.012) | (0.236, 0.328) | <0.001[*] |
| Same instructor for both treatment and control groups | 4 | 0.171 (0.234) | (-0.574, 0.915) | 0.518[+] |

*Note.* K = number of effect sizes, [*] indicates fixed ITS effects not included due to every study having a different ITS, [+] indicates that no ITS fixed effects were not included due to having a single ITS.

### 4.4.4. O Dimension

We coded four variables under the O dimension: type of measurement, subject of learning outcome, the type of question format, and measurement timing. However, the type of question format had a high proportion of missing values and was excluded. Table 7 presents effect sizes for each moderator in the O dimension. The analysis revealed a non-significant larger effect size with researcher-developed tests ($g=0.483$). The effect size with standardized tests was smaller ($g=0.200$) and statistically significant. The effect size for reading outcomes ($g=0.323$) was larger than the effect size for mathematics outcomes ($g=0.253$), and both were statistically significant. Outcomes measured immediately after the intervention produced a larger and effect size ($g=0.460$) than outcomes measured at end of the school year ($g=0.155$). Both effects were statistically significant.

**Table 7.** ITS Intervention Effect Sizes for O Dimension

| Moderator | K | g (SE) | 95% CI | p-value |
|---|---|---|---|---|
| **Type of Measurement** | | | | |
| Standardized test | 61 | 0.200 (0.009) | (0.166, 0.234) | 0.001 |
| Researcher-developed test | 16 | 0.483 (0.213) | (-0.154, 1.120) | 0.098* |
| **Type of learning outcome** | | | | |
| Mathematics | 22 | 0.253 (0.007) | (0.224, 0.281) | <0.001 |
| Reading and writing | 55 | 0.323 (0.039) | (0.218, 0428) | <0.001 |
| **Measurement Timing** | | | | |
| Immediately after the intervention | 23 | 0.460 (0.046) | (0.244, 0.677) | 0.013 |
| End of School Year | 47 | 0.155 (0.010) | (0.125, 0.185) | <0.001 |

*Note.* There was 1 study with measurement at the end of the unit, and 6 studies did not report measurement timing, * indicates fixed ITS effects not included due to every study having a different ITS.

### 4.4.5. S Dimension

We present the effect sizes by category of the S dimention in Table 8. Studies including urban ($g$=0.298) and suburban ($g$=0.260) settings had similar effect sizes. In contrast, studies including rural locations ($g$=0.146) had lower effect sizes. All three effect sizes were statistically significant.

**Table 8**. ITS Intervention Effect Sizes for S Dimension

| Moderator | K | $g$ (SE) | 95% CI | $p$-value |
|---|---|---|---|---|
| **Location** | | | | |
| Urban | 51 | 0.298 (0.010) | (0.259, 0.336) | <0.001 |
| Suburban | 45 | 0.260 (0.010) | (0.169, 0.350) | 0.015 |
| Rural | 43 | 0.146 (0.012) | (0.053, 0.239) | 0.029 |

Figure 2

### 4.4.6. MetaForest Results

We fit the MetaForest model 10 times using different seed values to ensure the stability of results. Each iteration converged with approximately 10,000 trees. We retained five moderators with permutation importance above the 50th percentile: dummy indicator for worked-out example, dummy indicator for type of learning outcome (i.e., mathematics), intervention duration, intervention condition, and dummy indicator for immediate measurement.

As the $R^2$ values varied across the 10 runs (times) of the MetaForest model, we report the median $R^2$ with the five retained predictors. The median out-of-bag $R^2$ was 0.09, and the median cross-validated $R^2$ was 0.25. The ranking of the five moderators remained stable across the 10 runs of the MetaForest algorithm. Based on permutation importance, the indicator of the ITS providing worked-out examples was ranked 1st as the most influential moderator. The dummy indicator of math ITS, intervention duration, intervention condition, and immediate measurement were ranked 2nd, 3rd, 4th, and 5th, respectively.

Figure 2 illustrates the marginal means of effect sizes by levels of the five important moderators, averaging all values of the other moderators. The plot displays point estimates (i.e., diamonds) with corresponding confidence intervals, offering insight into how effect sizes vary across different levels of each categorical moderator. The presence of worked-out examples is associated with slightly higher effect sizes compared to their absence, suggesting a positive contribution to intervention outcomes. Also, reading and writing ITS showed a higher effect size than math ITS. Intervention duration reveals a clear trend: effect sizes decrease as the duration of the intervention increases. Interventions lasting longer than six months show the smallest effect. Among the different intervention conditions, using the intervention as the main instructional method is associated with the highest effect size, whereas other categories (i.e., after-school programs, separate activity, homework) show smaller effects. Finally, immediate measurement is linked to slightly higher effect sizes than delayed measurement.

_______________________________________________

INSERT FIGURE 2 HERE

_______________________________________________

**4.4.7. Publication bias**

     Figure 3 is the Funnel plot of the standard error against observed effect sizes (k = 77) from the meta-analysis model without accounting for the fixed effect of ITS names. The diagonal lines represent the expected 95% confidence limits. Effect sizes from larger studies are toward the top, while those from smaller studies are more widely dispersed toward the bottom. There is a cluster of studies with more precise estimates and with effect sizes above zero but below 0.5. However, the spread of studies outside the funnel indicates heterogeneity of treatment effects beyond sampling error.

Figure 2. Marginal Mean Effect Sizes by Levels of Important Moderators

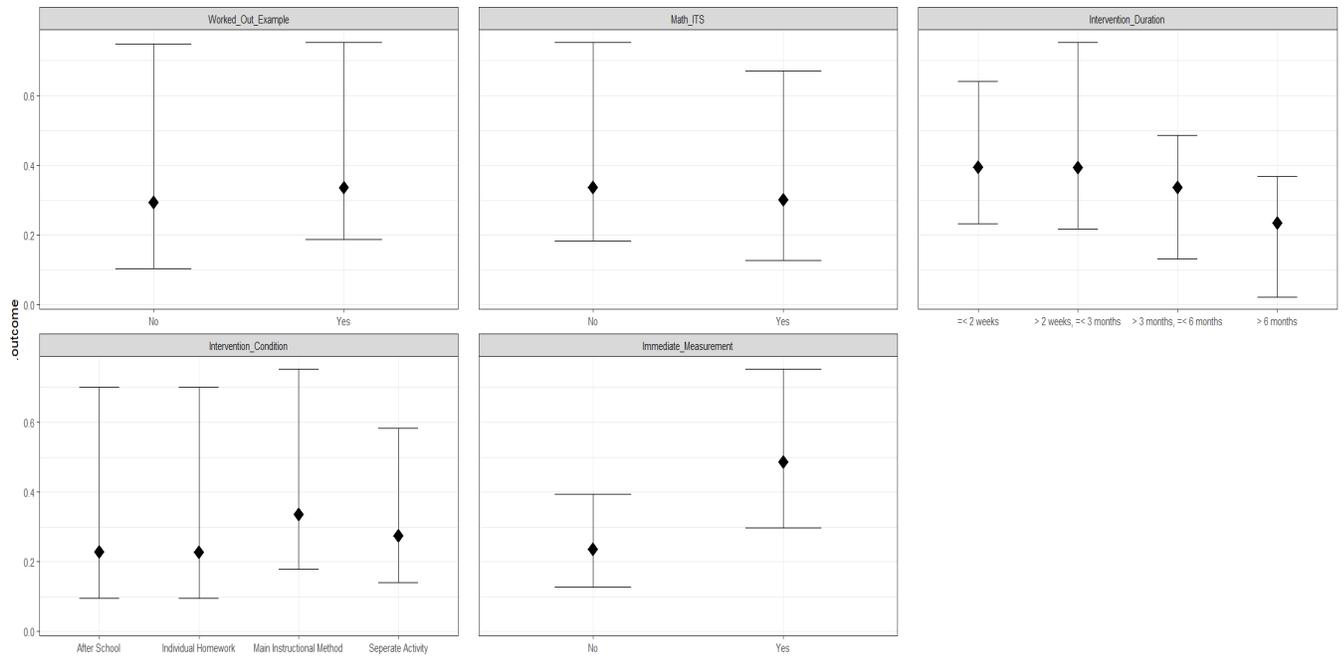

In this paper, we calculated fail-safe N with the Rosenthal approach (Rosenthal, 1979) and Orwin approach (Orwin, 1983). For Rosenthal approach, the fail-safe N is 36210 with target significance level of 0.05, while the fail-safe N is 191 for Orwin approach with target effect size of 0.10.

Figure 3. Funnel Plot of All Effect Sizes

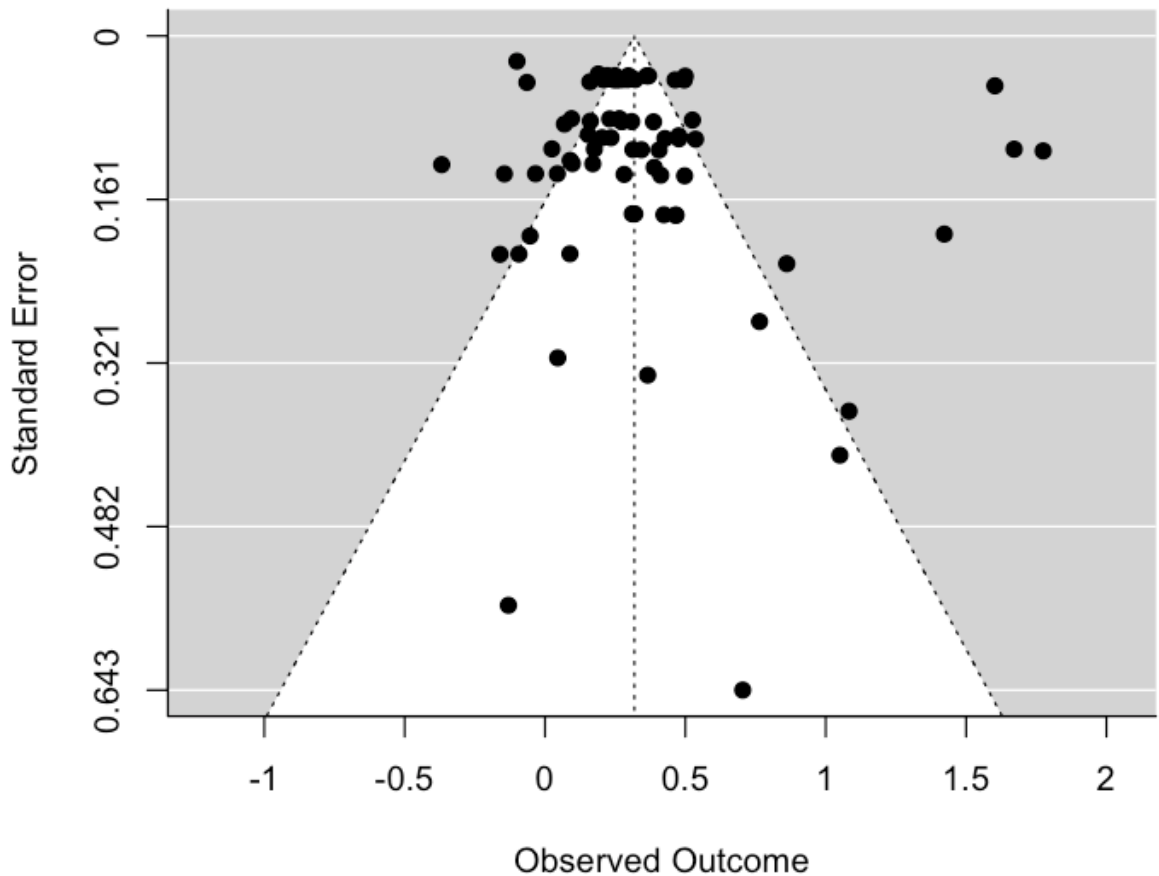

## 5. Discussion

In a recent survey of technology in the schools, the National Center for Education Statistics estimated that 45% of U.S. schools have a computer for each student, and 37% have a computer for each student in some grades or classrooms (Machusky & Herbert-Berger, 2022). Furthermore, 55% of schools reported using self-contained instructional packages to a moderate or large extent, which include ITS. However, there is no specific estimate of the availability of ITS in U.S. schools. Given the large availability of computers in U.S. schools, there is strong potential for the growth of ITS use to support classroom instruction, towards the availability and use of ITS in K-12 education at scale. Therefore, it is important to understand in which conditions the use of ITS is effective, so that resources can be allocated appropriately to maximize their benefit.

We found support for our Hypothesis 1, because our estimate of the effect of ITS was positive and statistically significant. Previous meta-analyses also found positive effects of ITS (Kulik & Fletcher, 2016; Ma et al., 2014; Steenbergen-Hu & Cooper, 2013, 2014; VanLehn, 2011), but effect size estimated in the current study (i.e., g = 0.271) was smaller than those of most previous meta-analyses. However, none of the previous meta-analyses targeted studies in U.S. K-12 schools that met the WWC standards without reservations. We found that only 8 of the 18 studies reviewed reported differential attrition. From these, five meet the WWC standards without reservation because they had low attrition, while the other three meet the standards with reservations. Whether WWC standards are met in the 10 other studies is uncertain, as they did not report attrition information.

Only one of the previous meta-analyses examined moderation by attrition level (i.e., Ma et al., 2014), reporting the largest effect size for studies not reporting attrition ($g$=0.48), followed by no attrition ($g$=0.39) and some attrition ($g$=0.29). We found that the effect sizes for studies with low or high attrition were not statistically significant, but the magnitude of the effect sizes was similar. Thus, our results do not support our Hypothesis 2, in contrast with Ma et al.'s results. We found a significant effect size for studies not reporting attrition ($g$=0.251), but lower than those reported by Ma et al. (2014). This difference between our results and Ma et al.'s (2014) may be because we only included experimental studies, while Ma et al. (2014) also included quasi-experimental and observational studies. For these design types, the estimates of treatment effect may be affected by both selection bias and attrition bias. These contrasting results indicate a limitation of the internal validity evidence provided by meta-analyses of ITS.

We found differences in effect sizes depending on the level of random assignment, but the effect sizes by levels of clustering and type of design (i.e., simple or blocked) were similar. None of the previous meta-analyses looked at these design characteristics. The level of random assignment is important in determining the potential for bias due to spill-over effects (Hong, 2015), but it is unclear from the included studies how much spill-over effects are a threat. We found that the most common design for evaluating ITS is a multisite cluster randomized trial with three levels of clustering and assignment at level 2. These designs are less vulnerable to spill-over effects than designs with level 1 assignment, but more vulnerable than those with level 3 assignment. Although the effect size of studies with level 3 assignment was not statistically significant, a larger effect for studies with assignment at the school level may be explained by higher levels of teacher collaboration and support that is facilitated when there is school-wide implementation of a program. Therefore, compliance with study specifications may be higher when all classrooms within a school are using the ITS. Although blocking has the potential to increase the statistical power of experimental evaluations of ITS, we found it was underutilized.

For evidence of the external validity of ITS, we discuss the results for the U, T, O, and S dimensions in this order. In the U dimension, ITS used in elementary or middle schools had similar effects. We also found statistically significant effects of ITS used with and without targeting low-achieving students that were close in magnitude. This contrasts with Steenbergen-Hu and Cooper (2013), who found that the benefits of ITS were larger for general-education students than for low-achieving students. While their results raise concerns that ITS could unintentionally widen achievement gaps, our study did not replicate their findings. Instead, our results support the wide implementation of ITS in elementary and middle schools regardless of students' previous achievement levels, which is aligned with the goal of ITS to provide personalized learning opportunities that benefit all students.

We found support for Hypothesis 3, because the MetaForest results indicated that among the five most important moderators of effect sizes, three were in the T dimension (i.e., worked-out example, intervention duration, and intervention condition), while two were in the O dimension (i.e., type of learning outcome and immediate measurement). Whereas earlier meta-analyses (e.g., Steenbergen-Hu & Cooper, 2013, 2014; Ma et al., 2014) identified general benefits of ITS without isolating the impact of a particular type of support, the current meta-analysis extends prior work by offering a more granular examination of the effect sizes of specific support types within ITS. Ma et al. (2014) compared ITS that provided feedback with those that did not, but Ma et al. (2014) did not elaborate on whether the feedback was in the form of a hint, explanation, or solution. We compared six different types of support provided by ITS (see Table 6), showing significant effect sizes of ITS providing on-demand hints, motivation support, worked-out examples, and reading materials.

Only the worked-out example feature had permutation importance above the $50^{th}$ percentile, indicating it was the most relevant type of support in predicting the ITS effect size. This result highlights the value of embedding structured problem-solving demonstrations within ITS environments. This on-demand feature allows students to choose when and how much to consult worked-out examples. This finding, taken together with the statistically significant effect of on-demand hints, indicates that ITSs that offer student choice succeed in enhancing learning.

In the current study, the effect of just-in-time hints was not statistically significant. Just-in-time hints, which are automatically triggered by the system based on student behavior or errors, have been critiqued for potentially interrupting students' cognitive processes and reducing opportunities for productive struggle (Roll et al., 2011) and reducing deeper mathematical thinking. In contrast, on-demand hints have been associated with more reliable learning gains, as they allow learners to regulate the timing and context of support (Vanacore et al., 2024). However, the effectiveness of on-demand features of ITS, such as worked-out examples and hints, may be moderated by students' self-regulated learning abilities. While more self-regulated learners can use on-demand ITS features strategically to deepen understanding, less self-regulated learners may exploit these features to game the system, rapidly requesting hints or answers without engaging meaningfully with the material (Aleven et al., 2016; Aleven et al., 2003; Baker et al., 2004; Wittwer & Renkl, 2008). Therefore, the design of ITS should consider not only the timing of support but also how to scaffold and encourage productive help-seeking behaviors.

In the T dimension, we found significant effects of employing ITS as the main instructional method and for individual homework. These effect sizes were estimated with regular classrooms as the control group, contrasting with Van Lehn's (2011) study, which used one-to-one human tutoring as the control group. Comparison with regular classrooms is critical because this setting is available to all students, and one-to-one human tutoring is costly and sparsely available. Our results contrast with those from Ma et al.'s (2014) meta-analysis, which showed stronger effects when ITS were used as a supplemental in-class activity, supplemental after-class instruction, or homework than as principal instruction or integrated class instruction.

We found a significant effect size for studies with a duration between three months and six months, but not for shorter or longer studies. However, the estimated effect in studies with a duration between 2 weeks and three months was much larger than in longer studies, but the variance was also large. This large variance may be because some short studies may not allow enough time for students to become familiar with the ITS, resulting in deflated effect sizes. In contrast, in other short studies, the effects may be inflated by novelty and over-alignment with specific learning goals. Effect sizes for studies longer than six months may have been affected by student fatigue and disinterest in using the ITS. Steenbergen-Hu and Cooper (2013) and Ma et al. (2014) also found smaller effects for studies lasting more than one year. The current results raise considerations for at-scale implementation of ITS in K-12 settings, where these systems are available to students throughout the academic year. To prevent the decline of engagement with ITS, teachers should consider classroom orchestration strategies (Dillenbourg, 2013; Prieto et al., 2015) for ITS that insert them into focused, spaced modules rather than continuous usage throughout the week. Also, ITS developers may need to incorporate mechanisms that sustain learner engagement and adapt dynamically to students' changing needs over time. This could include features such as personalized learning trajectories, embedded goal-setting tools, or integration with student-led individual and group projects.

In the O dimension, we found a significant effect size for outcomes measured by standardized tests, and a much larger but non-significant estimate with non-standardized tests. Xu et al. (2019) estimated an effect size with researcher-designed tests that was more than four times the size of the effect size with standardized tests. However, their estimate ($g = 0.25$) with standardized tests was similar to the one in the current study ($g = 0.20$). This contrasts with Ma et al.'s (2014) finding of similar effect sizes with both standardized and non-standardized effects estimated with a random effects model. Although there is disagreement across meta-analyses about the difference in effect sizes by type of measurement, estimates from researcher-developed measures are vulnerable to threats to construct validity (Shadish et al., 2002). More specifically, researcher-developed measures may measure specific facets of a construct that are closely aligned with ITS features, rather than all facets of a construct (Huggins-Manley et al., 2019).

In the S dimension, we found significant effect sizes of ITS in urban, suburban, and rural settings. However, it was noticeable that the effect size for studies that included rural settings was about half of the effect size of those implemented only in urban and suburban settings. None of the previous meta-analyses explicitly examined or reported differences in ITS effectiveness across urbanicity settings of K-12 students. Previous research has shown that rural schools are more likely to have infrastructure limitations (i.e., outdated computers, slower internet speeds) that prevent reliable implementation of ITS, higher student-to-computer ratios, limited computer use proficiency among teachers, and low access to teacher professional development for technology (Kim & Wargo, 2025; Mustafa et al., 2024). These factors help explain the lower effect size of ITS in studies that included rural schools.

The ITS with the most effect sizes included in this meta-analysis was the ITSS, with 39 effect sizes from five studies published across 10 years (i.e., Wijekumar et al., 2022; Wijekumar et al., 2012). The effect size of ITSS was $g = 0.472$ (SE =0.114), which was statistically significant ($p = 0.014$). The ITSS aims to improve reading comprehension by focusing on text coherence and signaling words. Because a substantial number of effect sizes were available for the ITSS, we evaluated the heterogeneity of effects for specific characteristics that varied across effect sizes. The effect of ITSS was larger with elementary school students ($g = 0.464$, SE = 0.085, $p < .001$) than with middle school students ($g = 0.395$, SE = 0.103, $p = 0.002$). The effect was larger when the study targeted a general population of students ($g = 0.500$, SE = 0.056, $p < .001$) than lower-achieving students (g = 0.278, SE = 0.037, p < .001). When the outcome was measured right after intervention, the effect size ($g = 0.752$, SE = 0.196, $p = .004$) was larger than when measured at the end of the academic year ($g = 0.343$, SE = 0.049, $p < .001$).

Among the ITS evaluated in this meta-analysis, ALEKS (Fang et al., 2018) and MATHia (Ritter & Fancsali, 2016) have been widely adopted in U.S. schools and offer valuable insight into the potential and challenges of scaling ITS in real-world educational settings. Interestingly, their effectiveness varied considerably, with ALEKS showing a negligible effect size ($g = 0.014$, SE = 0.112, $p = 0.908$), while MATHia demonstrated a much larger effect ($g = 0.544$, SE = 0.437, $p = 0.431$). Neither effect was statistically significant, but there were only six effect sizes available for each ITS, limiting statistical power. These are interesting examples of ITS because they have distinct strategies for implementation at scale. While MATHia has achieved scale with deep integration into classroom mathematics instruction, often supported by teacher professional development, ALEKS is more often used as a supplemental or independent tool, potentially limiting its instructional impact. Fang et al. (2018) performed a meta-analysis of experimental and quasi-experimental studies using ALEKS with a variety of control groups. For studies using a traditional classroom as a control group, they reported non-significant effect sizes with a fixed-effects model ($g = 0.04$, CI = [-0.05, 0.11]) and a random effects model ($g = 0.01$, CI = [-0.16, 0.18]). These cases highlight that reaching scale is not synonymous with effectiveness and underscore the importance of both pedagogical design and implementation context in realizing the potential of ITS at scale.

**6. Conclusion**

In this meta-analysis, we included 18 experimental studies evaluating 11 different ITS in K-12 U.S. schools. We used the PRISMA standards for reporting (Page et al., 2021). We focused on rigorous studies that would meet WWC standards without reservations. However, a few studies had high differential attrition, which downgrades them to meet the standards with reservations. Furthermore, another set of studies did not report differential attrition, which makes it impossible to know if they would meet the standards with or without reservations. Our selection criteria allowed rigorous quasi-experimental designs, which meet the WWC standards with reservations, but none were identified. This indicates an under-utilization of rigorous quasi-experimental designs, such as regression discontinuity designs and propensity score matched designs, for ITS evaluation.

One limitation of the current meta-analysis is that a few ITS are represented by only a single study, which restricts the ability to examine between-intervention variability or conduct meaningful subgroup analyses. Additionally, some potential moderators were either invariant across studies or had substantial missing data. Within the M dimension, our search identified experimental designs but no quasi-experimental designs, which made it impossible to evaluate whether the type of design is a moderator. The type of control group was also constant, only including classroom instruction.

External validity evidence in the U dimension was limited because student gender distribution and student ethnicity were not only frequently unreported but also inconsistently reported, limiting our ability to assess their moderating effects. The grade-level distribution of participants was also uneven, with only two studies focused on high school students, while the remainder targeted elementary and middle school populations, thus limiting generalizability to older students. Conclusions related to the T dimension were limited because classroom size was frequently not reported by studies. Within the T dimension, each ITS can be viewed as a treatment variation because they have a unique combination of supports. However, the frequency of the type of support (e.g., worked-out examples, hints) was not balanced in the dataset, which affects the power to detect moderation by specific supports. This imbalance is expected because some supports (e.g., hints) are more common in ITS for mathematics than for reading and writing.

The findings of this meta-analysis underscore a significant limitation in the current body of research on ITS: the lack of consistent, comprehensive, and transparent reporting in primary studies. To strengthen the evidence, base for the generalizability of results (i.e., external validity), transparent reporting of study design, participant characteristics, implementation contexts, and outcome data is essential. Thus, future experimental and quasi-experimental evaluations of ITS should adhere to established reporting standards, such as those Standards of the What Works Clearinghouse (U.S. Department of Education et al., 2022), the Standards for Excellence in Education Research (SEER; (Institute of Education Sciences, 2025), and APA's Journal Article Reporting Standards (American Psychological Association, 2025).

The current meta-analysis indicates a clear need for research on long-term use of ITS at scale in K-12 school settings. From the effect sizes included in this meta-analysis, only 20% were drawn from studies examining ITS implementations lasting more than six months. As the availability of ITS increases and they become more integrated into classroom instruction and digital learning ecosystems, it is essential to understand not only their long-term effectiveness but also how their use evolves over time when available at scale. Longitudinal studies that examine issues such as student engagement over time, student self-regulated use of ITS features, teacher orchestration of ITS in the classroom, curriculum alignment, and differential impacts across diverse learner populations are critical. Also, future research should explore how ITS can be designed and implemented to maintain their instructional value throughout the school year, as well as the systemic support for ITS use, such as professional development, infrastructure, and policy frameworks. These studies will help bridge the gap between short-term experimental findings and the realities of long-term implementation in K-12 education.

VanLehn, K. (2011). The Relative Effectiveness of Human Tutoring, Intelligent Tutoring Systems, and Other Tutoring Systems. *Educational Psychologist*, *46*(4), 197–221. https://doi.org/10.1080/00461520.2011.611369

Veritas Health Innovation. (2023). *Covidence systematic review software,*. In www.covidence.org

Viechtbauer, W. (2010). Conducting meta-analyses in R with the metafor package. *Journal of Statistical Software*, *36*(3), 1–48.

Wijekumar, K., Meyer, B. J., Lei, P., Beerwinkle, A. L., & Joshi, M. (2020). Supplementing teacher knowledge using web-based Intelligent Tutoring System for the Text Structure Strategy to improve content area reading comprehension with fourth- and fifth-grade struggling readers. *Dyslexia*, *26*(2), 120–136. https://doi.org/10.1002/dys.1634

Wijekumar, K., Meyer, B. J. F., & Lei, P. (2017). Web-Based Text Structure Strategy Instruction Improves Seventh Graders' Content Area Reading Comprehension. *Journal of Educational Psychology*, *109*(6), 741–760. https://doi.org/10.1037/edu0000168

Wijekumar, K., Meyer, B. J. F., Lei, P. W., Hernandez, A. C., & August, D. L. (2018). Improving content area reading comprehension of Spanish speaking English learners in Grades 4 and 5 using web-based text structure instruction. *Reading and Writing*, *31*(9), 1969–1996. https://doi.org/10.1007/s11145-017-9802-9

Wijekumar, K., Meyer, B. J. F., Lei, P. W., Lin, Y. C., Johnson, L. A., Spielvogel, J. A., & Shurmatz, K. M. (2014). Multisite Randomized Controlled Trial Examining Intelligent Tutoring of Structure Strategy for Fifth-Grade Readers. *Journal of Research on Educational Effectiveness*, *7*(4), 331–357. https://doi.org/10.1080/19345747.2013.85333352

# Appendix 1. Search Terms

(shown here in the will the ACM digital library format): [[Abstract: "intelligent tutor*"] OR [Abstract: "artificial tutor*"] OR [Abstract: "computer tutor*"] OR [Abstract: "computer-assisted tutor*"] OR [Abstract: "computer-based tutor*"] OR [Abstract: "intelligent learning environment*"] OR [Abstract: "computer coach*"] OR [Abstract: "online tutor*"] OR [Abstract: "e-tutor*"] OR [Abstract: "electronic tutor*"] OR [Abstract: "web-based tutor*"] OR [Abstract: "intelligent virtual"] OR [Abstract: "intelligent agent"] OR [Abstract: "cognit* tutor*"] OR [Abstract: "adapt* tutor*"] OR [Abstract: "virtual companion"] OR [Abstract: "intelligent coach*"]] AND [Abstract: student*] AND NOT [Abstract: college] AND NOT [Abstract: undergraduate] AND [Publication Date: (01/01/2011 TO 12/31/2021)]

**Appendix 2. Study Inclusion criteria**

1. One of the systems examined in the study meets definition of intelligent tutoring system: use Artificial Intelligence (AI) techniques to adapt and scaffold the experience of the individual learner in ways that attempt to maximize the quality of their learning and/or minimize their learning time.
2. Experimental study or propensity score analysis (matching, weighting, stratification) study or regression discontinuity design of intelligent tutoring systems. These are the designs that are accepted by the What Works Clearinghouse to meet standards of evidence. For propensity score analysis designs, baseline equivalence must be demonstrated.
3. Studies were published between January 1$^{st}$ 2011 and December 31$^{st}$ 2021.
4. Studies had to focus on students in grades K–12.
5. Studies had to measure the effectiveness of ITS on student achievement (e.g. tests developed by researchers, standardized tests, high-stakes tests).
6. Studies had to have used an independent comparison group that was non-ITS.
7. Studies had to be conducted with a sample from the United States of America.
8. Studies published in academic journals, dissertations/theses and conference proceedings.

**Study Exclusion Criteria**

1. Studies focusing exclusively on students with learning disabilities or social or social or emotional disorders (e.g., students with attention-deficit/hyperactivity disorder)
2. Studies where information about effect sizes could not be obtained (either from the paper, online resources, or by contacting the authors).

3. Studies where no information about sample demographics (grade level, gender, minority status) could be obtained (either from the paper, online resources, or by contacting the authors) were excluded.

4. Studies without a comparison group or those with one-group pretest–posttest designs were excluded.

5. Studies just comparing different ITS and variations of an ITS were excluded.

6. None of the outcomes is student achievement

**Appendix 3. Study Characteristics Coded with MUTOS Framework**

| MUTOS Dimension | Study Characteristics | Example Values |
|---|---|---|
| Methods | Type of Publication | Journal Article, Conference Proceeding, Dissertation or Unpublished |
| | Type of Study Design | Experimental Design, Propensity Score Analysis, or Regression Discontinuity Design |
| | Assignment Level* | Level 1, 2, or 3 |
| | Levels of clustering* | 2 or 3 |
| | Simple or blocked design* | Simple or blocked |
| | Attrition | Number of participants that dropped out of the study |
| Units | Grade Level | Elementary, Middle, High |
| | Gender | Male Or Female Percentage |
| | Ethnicity | Asian, Black, Hispanic or White, |
| | Prior Domain Knowledge | Low, Medium, High, Varied or Not Reported |
|  | Students' Disadvantage Targeting | Low Achiever, Students in Poverty, English Language Learner or No targeting |
| Treatments | ITS used | ITSS, MATHia, ALEKS |
|  | Intervention Support Type | Question Sequence, Video Sequence, Reading Materials, Feedback, Hints, Other, or Not Reported |
|  | Number Of Intervention Groups | Specific Number |
|  | Number Of Control Groups | Specific Number |
|  | Duration Of Intervention | Time Length in hours, days, weeks and months |
|  | ITS Intervention Condition | ITS As Principal Instruction, ITS-Integrated In-Class Activities, ITS-Supplemented In-Class Activities, ITS-Supplementary After-Class Instruction, ITS-Assisted Homework, Other |
| Operations | Type of Measure | Research Developed, Standardized Test, |
|  | Subject of Learning Outcome | Mathematics, Writing, Reading, Science, Social studies or Other |

| | The Type of Question Format | Multiple Choice, Short Answer, Essay and Other |
| | Measurement Timing | End of School Year, End of Semester, Immediately After Intervention, Not reported |
| | Instructor Effects | Different Instructors: Different Teachers Taught Treatment and Control Groups, Same Instructor: Same Teacher or Teachers Taught Treatment And Comparison Groups, Not reported |
| Settings | | |
| | Location Type of Participants | Rural, Urban, Suburban, Not reported |

**Note.** The initial coding classified the assignment level into student, classroom, teacher, and school categories. Because a clear distinction between teacher and classroom random assignment was not possible, the studies were recoded by the first and second authors using the variables marked with a *.

**Appendix 4. Standardized Mean Differences (SMD)**

The equation of SMD is defined as (Borenstein et al., 2009):

$$SMD = \frac{M_1 - M_2}{SD_{pooled}} \quad (1)$$

Where $M_1$ and $M_2$ are the mean values of intervention and control group respectively. The denominator is the pooled standard deviation across two groups, which is defined as (Borenstein et al., 2009):

$$SD_{pooled} = \sqrt{\frac{(n_1-1)S_1^2 + (n_2-1)S_2^2}{n_1+n_2-2}} \quad (2)$$

Where $n_1$ and $n_2$ are the sample size of treatment and control groups respectively. $S_1$ and $S_2$ are the standard deviations in those two groups.

When applying WLS, the estimator of variance of regression coefficients could be defined as (Tipton, 2015):

$$V(\beta) = \left(\sum_{k=1}^{K} X_k' W_k X_k\right)^{-1} \left(\sum_{k=1}^{K} X_k' W_k \Sigma_k W_k X_k\right) \left(\sum_{k=1}^{K} X_k' W_k X_k\right)^{-1} \quad (9)$$

Where $X$ is a vector of covariates, $W$ is a diagonal weight matrix and $\Sigma$ is a covariance matrix of effect sizes for the $k^{th}$ study. If the effect sizes within the same study are statistically independent, the Equation 9 could reduce to $V(\beta) = \left(\sum_{k=1}^{K} X_k' W_k X_k\right)^{-1}$ since covariances of effect sizes will be zero (i.e., $\Sigma_k = 0$). However, the independence assumption cannot be met because there are multiple effect sizes within each study. Thus, the standard errors of regression coefficients might be biased, which will cause biased Type I error (Tanner-Smith et al., 2016; Tipton et al, 2019).

With RVE, the variance of regression coefficients could be given by (Tipton, 2015):

$$V^R(\beta) = (\sum_{k=1}^{K} X_k' W_k X_k)^{-1} (\sum_{k=1}^{K} X_k' W_k e_k e_k' W_k X_k)(\sum_{k=1}^{K} X_k' W_k X_k)^{-1} \quad (10)$$

Where $e_k$ is the residual variance of effect sizes cannot be explained by covariates.

**Appendix 5. Details of MetaForest Specification**

To construct moderators representing intervention support types, all categories of support type were converted to dummy indicators, except for feedback/explanations and question sequencing, because all 11 ITS had these two support types. The categorical moderators, student disadvantage targeting and location, were transformed into multiple dummy indicators. The categorical variable measuring whether the study targeted disadvantaged students was recoded into three dummy indicators: English learners, students in poverty, and low achievers. Student location was dummy-coded into urban, suburban, and rural dummy indicators. However, to reduce multicollinearity between these variables, only dummy indicators of studies targeting low achievers and rural student location were retained.

We fit the initial MetaForest model with a large number of trees (i.e., 20,000) to check for convergence. After confirming convergence, we reduced the number of trees to improve processing speed in the final model. Next, we used permutation importance (Altmann et al., 2010) to obtain importance values for all moderators. Then, we only retained moderators with importance value above the 50th percentile of permutation importance. We implemented random forests and permutation importance calculation with the R package *metaforest* (Van Lissa, 2024). With the reduced set of moderators, we used the R package *caret* (Kuhn, 2008) to perform 5-fold cross-validation, optimizing MetaForest hyperparameters (i.e., the number of variables per tree, and the minimum node size) based on root mean squared prediction error (RMSE).

The hyperparameter tuning process was conducted using a grid search to optimize model performance. The tuning grid included three weighting methods ("random," "fixed," and "unif"), the number of candidate variables per tree ranging from 2 to 5, and the minimum node size (i.e., min.node.size), varying from 2 to 10. The model evaluated different combinations of these parameters to identify the optimal configuration based on predictive accuracy. This process ensured that the final model was selected from a comprehensive range of potential hyperparameter values, balancing model complexity and performance. With the optimal tuning parameters and selected moderators, we ran the final model to rank the moderators.